\documentclass[10pt,twocolumn]{article}

\usepackage{graphicx}
\usepackage{amsfonts}
\usepackage{amsmath}
\usepackage{amssymb}
\usepackage{mathrsfs} 
\usepackage{lipsum}
\usepackage{color}
\usepackage[squaren,Gray,cdot]{SIunits}
\usepackage{authblk}
\usepackage[margin=2cm]{geometry}
\usepackage{mathtools, cuted}
\usepackage{array,amsmath,booktabs}
\usepackage[font=small, labelfont={bf,normalsize}]{caption}
\usepackage[version=3]{mhchem}

\usepackage{sectsty}
\sectionfont{\sffamily}
\subsectionfont{\sffamily}
\subsubsectionfont{\sffamily}
\paragraphfont{\sffamily}

\title{\sf Wettability patterning in microfluidic devices using thermally-enhanced hydrophobic recovery of PDMS}

\author[1]{Marc Pascual}
\author[1]{Margaux Kerdraon}
\author[1]{Quentin Rezard}
\author[1,2]{Marie-Caroline Jullien}
\author[1,3]{Lor\`{e}ne Champougny}

\affil[1]{\normalsize Gulliver, CNRS, ESPCI Paris, PSL University, 10 rue Vauquelin, 75005 Paris, France}
\affil[2]{Institut de Physique de Rennes, UMR CNRS 6251, B\^{a}t. 11A, Campus de Beaulieu, 263 avenue du G\'{e}n\'{e}ral Leclerc, 35042 Rennes CEDEX, France}
\affil[3]{Grupo de Mec\'anica de Fluidos, Departamento de Ingenier\'ia T\'ermica y de Fluidos, Universidad Carlos III de Madrid, Av. Universidad 30, 28911 Legan\'es (Madrid), Spain}
\date{}
\begin{document}
%
%
\twocolumn[
\begin{@twocolumnfalse}
\maketitle
\begin{abstract}
Spatial control of wettability is key to many applications of microfluidic devices, ranging from double emulsion generation to localized cell adhesion. A number of techniques, often based on masking, have been developed to produce spatially-resolved wettability patterns at the surface of poly(dimethylsiloxane) (PDMS) elastomers. A major impediment they face is the natural hydrophobic recovery of PDMS: hydrophilized PDMS surfaces tend to return to hydrophobicity with time, mainly because of diffusion of low molecular weight silicone species to the surface. Instead of trying to avoid this phenomenon, we propose in this work to take advantage of hydrophobic recovery to modulate spatially the surface wettability of PDMS. Because temperature speeds up the rate of hydrophobic recovery, we show that space-resolved hydrophobic patterns can be produced by locally heating a plasma-hydrophilized PDMS surface with microresistors. Importantly, local wettability is quantified in microchannels using a fluorescent probe. This ``thermo-patterning'' technique provides a simple route to \textit{in situ} wettability patterning in closed PDMS chips, without requiring further surface chemistry. 
\end{abstract}
%
\vspace{0.5cm}
\end{@twocolumnfalse}]
%
\section{Introduction}
An increasing number of applications require hydrophilic / hydrophobic surface patterning, \textit{i.e.} space-dependent wettability properties on a single surface \cite{Xia2012,Ueda2013}. If these wettability patterns can be useful at macroscale, for instance in droplet deposition for printing techniques \cite{Tian2013} or control of heterogeneous nucleation of water \cite{Varanasi2009}, they find a considerable number of applications at microscale. These include pumpless fluid actuation \cite{Darhuber2005,Ghosh2014}, double emulsion generation \cite{Aserin2008}, wall-free flow control in open \cite{Oliveira2010} and closed \cite{Zhao2001,Wang2015} microfluidic devices, microdroplets generation and control \cite{Handique1997,Kobayashi2011} and biological cell patterning \cite{Tan2004,Patrito2007,Frimat2009}, to name a few. \medskip\\
\indent Poly(dimethylsiloxane) (PDMS), which is one of the most commonly used materials in the fabrication of microfluidic systems \cite{Xia1998}, offers naturally hydrophobic wetting conditions, therefore complicating the handling of aqueous solutions. 
As a consequence, a number of techniques have been developed to hydrophilize PDMS surfaces. 
They mostly follow two main routes: on the one hand, direct chemical modification of the surface through \ce{O2} plasma treatments \cite{Hollahan1970} or UV-ozone exposition \cite{Berdichevsky2004} or, on the other hand, surface coating with additional material layers. 
Physical techniques have also been proposed, involving for instance surface roughness modification \textit{via} laser ablation redeposits \cite{vanPelt2014} or oxygenated groups removal using physical contact treatment \cite{Guckenberger2012}. 

For the two main hydrophilization routes, a popular approach to achieve spatially-resolved wettability patterns on PDMS surfaces relies on masks or stencils to select the areas where the hydrophilizing treatment is applied.
For example, wettability patterns on PDMS can be obtained using a photomask \cite{Ma2011,Larson2013}, or simply epoxy glue \cite{Li2016} or marker \cite{Bodin2017} deposits, to shield portions of the surface from plasma treatment before chip sealing.
Some methods require more advanced chemistry, like selective graft polymerization of hydrophilic polymers, either initiated by UV exposure through a mask \cite{Hu2004,Patrito2006,Abate2008,Schneider2010} or inhibited by the $\ce{O2}$ contained in integrated gas reservoirs \cite{Romanowsky2010}.
Alternatively, mask-free techniques for wettability patterning include flow confinement of chemical treatment using an inert solution \cite{Abate2010} and spatially-controlled $\ce{O2}$ plasma oxidation by playing on the geometry of the PDMS chip and plasma parameters \cite{Kim2015}.\medskip \\
\indent Unless permanent grafting of hydrophilic polymers is achieved \cite{Hu2004,Patrito2006,Abate2008,Schneider2010,Romanowsky2010}, the hydrophobic recovery of PDMS surfaces, \textit{i.e.} their natural ability to restore their hydrophobicity after an oxidizing treatment (air/oxygen plasma \cite{Morra1990,Toth1994,Hillborg2001}, UV-ozone \cite{Hillborg2004,Ma2011} or electric discharge \cite{Hillborg1998,Kim2000}), is usually seen as a major impediment to wettability patterning. 
This phenomenon is attributed to a combination of different mechanisms, including (i) reorientation of surface polar groups towards the bulk and of non-polar groups towards the surface \cite{Morra1990,Toth1994}, (ii) elimination of hydroxyl groups by condensation of silanols \cite{Morra1990} (iii) diffusion of low molecular weight silicone species (either preexisting or produced \textit{in situ} by the oxidizing treatment) from the bulk to the surface \cite{Toth1994,Fritz1995,Kim2000,Kim2001}.
The latter mechanism (iii) is usually thought to provide the major contribution to hydrophobic recovery \cite{Toth1994,Kim2001}.
The rate of hydrophobic recovery can therefore be decreased by removing low molecular weight chains (through solvent extraction \cite{Toth1994} or extra curing time before plasma \cite{Eddington2005}) or by slowing down the diffusion of low molecular weight species (for example by storage at low temperature \cite{Bodin2017}).
Conversely, studies on PDMS macroscopic samples showed that hydrophobic recovery is accelerated when increasing temperature \cite{Morra1990,Hillborg2001,Kim2000,Chen2007}.
Additionally, oxidized PDMS surfaces have been observed to be covered with a thin brittle silica-like layer, in which cracks provide preferential channels for the diffusion of low molecular weight species towards the surface \cite{Owen1994,Fritz1995,Hillborg2001,Bodas2007,Befahy2010}. \medskip \\
In this work, instead of considering hydrophobic recovery as an impediment, we propose to take advantage of its temperature dependence to locally tune the wettability of PDMS surfaces.
The idea consists in performing local heating of a plasma-hydrophilized PDMS surface with a microresistor \cite{Miralles2015}, thus accelerating the hydrophobic recovery where a hydrophobic pattern is desired, while keeping the rest of the surface hydrophilic.
To our knowledge, such a space-dependent tuning of hydrophobic recovery for patterning purposes has never been reported in the literature.
We show that our method allows \textit{in-situ} patterning of sealed microchannels without need of chemically advanced surface treatments, and with a resolution of a few hundreds of microns.
In general, wettability properties are probed at macroscale using contact angle measurements.
Inspired by the work of Guckenberger \textit{et al.} \cite{Guckenberger2014} we develop and validate the use of a fluorescent protein (BSA-FITC) as a quantitative wettability probe at in microchannels.
Our experimental materials and procedures are described in section \ref{sec:exp}. In section \ref{sec:results}, we then present our results on the temperature dependence of hydrophobic recovery at macroscale (using contact angle measurements) and at microscale (using fluorescent protein adsorption as a local wettability probe), before finally demonstrating hydrophobicity patterning in microfluidic chips with our technique. 
%
\section{Materials and methods}\label{sec:exp}
%
\begin{figure}[t]
\centering
\includegraphics[width=\linewidth]{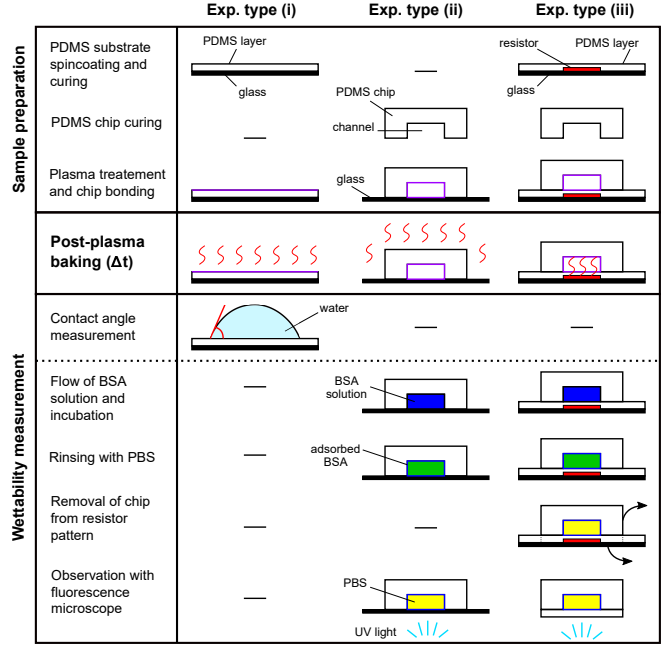}
\caption{Schematics of the experimental protocol implemented, with three major steps: sample preparation, hydrophobic recovery through post-plasma baking and wettability measurements. These steps are addressed differently, depending on the type of experiment performed: (i) global heating of macroscopic PDMS surfaces, (ii) global heating of PDMS microchannels and (iii) local heating of PDMS microchannels.}
\label{fig:protocole}
\end{figure}
Three experimental configurations are studied, going progressively from macroscale to microscale: (i) PDMS macroscopic surfaces submitted to a global heating, (ii) PDMS microchannels submitted to a global heating and (iii) PDMS microchannels submitted to a local heating.
As summarized in Fig.~\ref{fig:protocole}, all kinds of experiments are performed following the same succession of steps (although using different techniques).
First, solid PDMS surfaces are fabricated and made hydrophilic with $\ce{O2}$ plasma treatment (Sec. \ref{ssec:PDMS}).
These surfaces are then heated, either globally or locally (Sec. \ref{ssec:temp_control}), at various temperatures and for various durations $\Delta t$ to allow for hydrophobic recovery.
Finally, the wettability of the hydrophobically recovered surface is assessed (Sec. \ref{ssec:wettability}).
%
\subsection{PDMS surfaces and chip fabrication}\label{ssec:PDMS}
%
Throughout the whole study, PDMS elastomers are made of Sylgard 184 (Dow Corning) with a weight ratio 1/10 of crosslinker/polymer.
For macroscopic experiments (i), the sample consists of a $30\pm5~\micro\meter$-thick PDMS layer spin-coated onto a silicon wafer and subsequently cured at $70~\celsius$ for 2 hours.
For the purpose of microscale studies (ii) and (iii), a straight channel design ($400~\micro\meter$ in width, $2.5~\centi\meter$ in length and from $20$ to $30~\micro\meter$ in height) was created on a silicon wafer using standard soft lithography techniques \cite{Xia1998}. Liquid PDMS is poured onto this mold, cured at $70~\celsius$ for 2 hours; inlets and outlets of $0.75~\milli\meter$ in diameter are punched into the chips, which are finally bonded onto a substrate.
For experiments of type (ii) (global heating), this substrate is simply a $150~\micro\meter$-thick glass cover slip. For experiments of type (iii), it consists in a $30\pm 5~\micro\meter$-thick PDMS layer, cured at $70~\celsius$ for 2 hours, on top of the resistor pattern used for local heating (see Fig.~\ref{fig:resistors} and description in the section below).
The hydrophilization of all PDMS surfaces (and chip bonding in experiments (ii) and (iii)) is achieved with an $\ce{O2}$ plasma treatment (CUTE apparatus, Femtoscience). The plasma is generated by applying a power of $100~\watt$ (experiments (i)) or $25~\watt$ (experiments (ii) and (iii)) at a frequency of $50~\kilo\hertz$ during $50~\second$ within a vacuum chamber filled with oxygen at a pressure of 0.5 Torr (67 Pa). 
Note that the potential impact of using two slightly different hydrophilization protocoles for experiments of type (i) and (ii)-(iii) will be further discussed in paragraph \ref{ssec:results_micro}.
%
\subsection{Global and local temperature control}\label{ssec:temp_control}
%
\begin{figure}[t]
\centering
\includegraphics[width=1\linewidth]{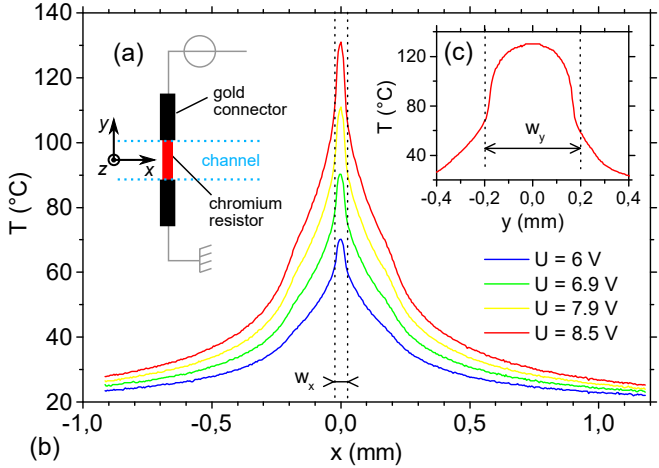}
\caption{(a) Sketch of the local heating device and its position with respect to the microfluidic channel to be patterned. --- (b) Temperature profiles along (b) the longitudinal $x$-axis and (c) the transverse $y$-axis at the surface of a $R = 0.78~\kilo\ohm$ micro-resistor of dimensions $w_x \times w_y = 50~\micro\meter \times 400~\micro\meter$. Tuning the applied voltage $U$ allows to change the peak temperature $T_{\mathrm{max}}$ of the profile.}
\label{fig:resistors}
\end{figure}
After hydrophilization, PDMS surfaces are baked at various temperatures and for various amounts of time to allow for hydrophobic recovery.
For experiments of type (i) and (ii), the samples are heated globally in conventional gravity ovens. In order to avoid contamination of PDMS surfaces in the case of experiments (i), the samples are stored in individual PELD or PYREX petri dish, depending on the temperature of the post-plasma baking (respectively under and above $70~\celsius$). 

For experiments of type (iii), a microscale heating system is fabricated from a $700~\micro\meter$-thick glass wafer covered with a $150~\nano\meter$ gold layer and a $15~\nano\meter$ chromium layer (A.C.M. France).
These layers are successively etched, using soft lithography printed S1818 photoresist (MicroChem) to protect the desired pattern of chromium resistors and gold connectors.
Fig.~\ref{fig:resistors}a shows the geometry of a resistor and its positioning with respect to the microchannel to be patterned.
Each chromium resistor is $w_x = 50~\micro\meter$ in width and $w_y = 400~\micro\meter$ in length.
Heat is produced by Joule effect when a voltage difference $U$ is applied to the resistor.

The temperature profile generated at the surface of a resistor (covered with a $30~\micro\meter$ insulating PDMS layer) is characterized with an infrared camera (FLIR - A6500SC) coupled to an infrared zoom lens (FLIR ATS). 
Importantly, the glass wafer is placed on a $2~\centi\meter$-thick aluminum block, serving as a heat sink to guarantee a stationary temperature profile despite the continuous application of heating power \cite{Selva2009}.
Fig.~\ref{fig:resistors}b shows the temperature profiles obtained along the longitudinal $x$-direction for different applied voltages $U$, yielding maximum profile temperatures $T_{\mathrm{max}} = 70, \, 90, \, 110 \text{ and } 130~\celsius$, respectively.
The inset (Fig.~\ref{fig:resistors}c) displays the temperature profile measured across the resistor in the transverse $y$-direction for $T_{\mathrm{max}} = 130~\celsius$. The temperature remains reasonably homogeneous (\textit{i.e.} between $110$ and $130~\degreecelsius$) within $y=\pm 150~\micro\meter$ from the resistor's center but drops when approaching the edges, close to the gold connectors. Additionally, COMSOL simulations of heat diffusion in our geometry show that the maximum temperature also remains acceptably uniform along the vertical $z$-direction, dropping from $130~\celsius$ at the bottom to about $110~\celsius$ at the top of a $30~\micro\meter$-height microchannel.
%
\subsection{Wettability measurements} \label{ssec:wettability}
%
The wettability of the thermally-aged PDMS surfaces is finally evaluated. For macroscopic experiments (i), contact angles measurements are performed, while a fluorescent probe is used in microscale experiments (ii) and (iii), as described below.
\paragraph*{Advancing contact angle.}
%
For experiments of type (i), we determine the advancing contact angle $\theta_a$ of ultrapure water (Milli-Q) on macroscopic PDMS surfaces by direct image analysis using a commercial solution (PSA30-KRUSS). The sample is first left to cool down for one minute after the end of thermal aging. A $5~\micro\liter$ droplet is then deposited onto the substrate and inflated with an additional $5~\micro\liter$ at a rate of $30~\micro\liter\per\minute$. The advancing contact angle $\theta_a$ is measured just before depinning of the contact line. Each measurement is repeated five times on two different samples made of different batches of Sylgard 184. Error bars in Fig.~\ref{fig:macro} represent the standard deviation of these measurements.
\paragraph*{Local wettability measurements.}
%
Because contact angle measurements can hardly be performed in sub-millimetric channels \cite{Zhu2010}, we turn to a fluorescence-based method to assess the wettability of PDMS surfaces in microscale experiments (ii) and (iii). Only a few indirect wettability measurements using fluorescent probes have been performed in microchannels so far. Guckenberger \textit{et al.} \cite{Guckenberger2014} have used the local depletion of Red Nile coated on polystyrene to quantify the penetration of $\ce{O2}$ plasma treatment in microchannels. On PDMS surfaces, the hydrophilic copolymer PLL-g-PEG has been employed by Bodin \textit{et al.} \cite{Bodin2017} to qualitatively visualize hydrophilic channel surfaces. 

In order to probe the wettability of PDMS within microchannels, we use Bovine Serum Albumine tagged with Fluorecein Isothiocyanate (BSA-FITC, Sigma-Aldrich), as this fluorescent protein is known to preferentially adsorb on hydrophobic surfaces \cite{Vlachopoulou2009}. Similar protocols are used when the microchannel is heated globally (experiments (ii)) or locally (experiments (iii)), as sketched in Fig.~\ref{fig:protocole}. 
The fluorescent protein solution is made by dissolving BSA-FITC in Phosphate Saline Buffer (PBS; one tablet in $200~\milli\liter$ of deionized water: 0.01 M phosphate buffer, 0.0027 M potassium chloride and 0.137 M sodium chloride - pH 7.4, Sigma-Aldrich) at a concentration of $0.1~\gram\per\liter$. This concentration is larger than the threshold concentration $0.02~\gram\per\liter$ above which the native PDMS surface becomes saturated with BSA molecules \cite{Lok1983b}. 

After the post-plasma baking step, the microfluidic chip described in section \ref{ssec:PDMS} is first left to cool down to ambient temperature. The BSA-FITC solution is flown in the microchannel for $2~\minute$ by imposing an overpressure of $75~\milli\bbar$ with a pressure controller (Fluigent - MFCS 4C). The BSA-filled channel is then left to incubate for $30~\minute$, either with or without any flow. The incubation time was chosen long enough to reach a plateau of BSA adsorption at a given concentration, but short enough not to make the whole process too cumbersome. It was also experimentally noticed that keeping the BSA solution flowing during the incubation step led to a more homogeneous BSA adsorption, maybe due to a reduced adsorption of aggregates. The non-adsorbed proteins are finally rinsed by flushing the channel with PBS for $5~\minute$ with an overpressure of $75~\milli\bbar$.  

In order to quantify the amount of BSA-FITC adsorbed at the walls, the PBS-filled microchannel is observed using a fluorescence microscope (Zeiss Observer.A1) with 5x or 20x dry objective (N.A. 0.12 or 0.5). The sample is illuminated with UV light (HBO 103W/2 - OSRAM lamp, filtered at $495~\nano\meter$) and the fluorescence signal is recorded at a wavelength of $517~\nano\meter$, with an exposure time of 1 second (Andor Zyla camera).
%
\section{Results and discussion} \label{sec:results}
\subsection{Temperature dependence of hydrophobic recovery} \label{ssec:results_macro}
%
\begin{figure}[t]
\centering
\includegraphics[width=\linewidth]{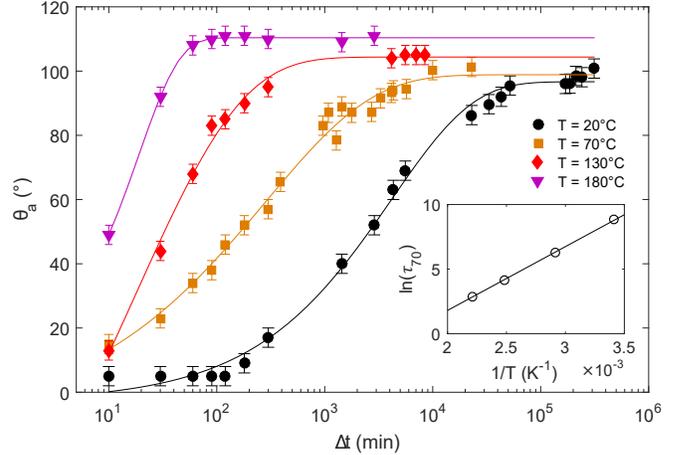}
\caption{Advancing contact angle $\theta_a$ of a sessile water drop on PDMS surfaces as a function of the post-plasma baking time $\Delta t$ and for various baking temperatures $T=20$ (room temperature), $70$, $130$ and $180~\celsius$. For each temperature, the solid line is a fit of the data (symbols) using a stretched exponential function. The error-bars correspond to the maximal dispersion of the data. --- Inset: Arrhenius plot of the recovery time $\tau_{70}$, needed for the PDMS surface to go back to $\theta_a = 70\degree$, \textit{i.e.} plot of the logarithm of $\tau_{70}$ (in minutes) as a function of the inverse of the temperature $T$. The solid line is a linear fit of the data (symbols).}
\label{fig:macro}
\end{figure}
Experiments of type (i) aim at quantifying how the hydrophobic recovery of macroscopic PDMS samples depends on temperature. The advancing contact angle $\theta_a$ of a sessile water drop deposited on the sample surface is measured for different baking times $\Delta t$ after the plasma treatment. Fig.~\ref{fig:macro} shows the time evolution of $\theta_a$ for PDMS samples baked at different temperatures: $T = 20$ (room temperature), $70$, $130$ or $180~\celsius$. The data show that the hydrophobic recovery of PDMS is complete after several months at room temperature, but can be achieved in only a few hours when the baking temperature is increased up to $180~\celsius$.
For interpolation purposes, the hydrophobic recovery curve at each temperature is fitted by a stretched exponential function of type $\theta_a = a + b \, \exp[-(\Delta t/c)^d]$, where $a$, $b$, $c$ and $d$ are adjustable parameters, as inspired by Kim \textit{et al.} \cite{Kim2000}. The characteristic time $\tau_{\Theta}$ is defined as the time needed, for a sample baked at a given temperature, to reach $\theta_a = \Theta$ (in degrees) and is deduced from the fitted functions (solid lines in Fig.~\ref{fig:macro}). As an example, the inset of Fig.~\ref{fig:macro} shows an Arrhenius plot of $\tau_{70}$, that is to say the evolution of $\ln (\tau_{70})$ with the inverse of the temperature $1/T$.

As already observed in other studies \cite{Morra1990,Kim2000,Hillborg1998,Hillborg2001}, the Arrhenius plot of $\tau_{70}$ is linear. Additionally, we find that the corresponding slope is essentially independent of the definition chosen for $\tau_{\Theta}$ for target angles in the range $\Theta = 60 - 90\degree$.
The characteristic time for hydrophobic recovery may thus be written as 
\begin{equation}
\tau_{70} \propto \exp (E_a /RT),
\label{eq:tau}
\end{equation}
where we introduce the universal gas constant $R = 8.314~\joule\per\mole\per\kelvin$ and an activation energy $E_a$. The activation energy $E_a = 42 \pm 4~\kilo\joule\per\mole$ deduced from the data in Fig.~\ref{fig:macro} is in good agreement with numerical values reported in the literature \cite{Morra1990,Kim2000,Hillborg1998,Hillborg2001}. Remarkably, these values all lie in the range $E_a = 30-60~\kilo\joule\per\mole$ for various hydrophilization processes (electrical discharges, air or oxygen plasma), exposure times and PDMS crosslink densities. 

The Arrhenius behavior of the hydrophobic recovery time and its robustness may be understood as follows. Diffusion of species in elastomers can usually be described as an activated process\cite{McKeen2016}, where the diffusion coefficient has the form $D \propto \exp (-E_a /RT)$. Then, assuming the hydrophobic recovery of PDMS is mainly controlled by the diffusion of low molecular weight species to the surface, the recovery timescale $\tau_{70}$ and the diffusion coefficient $D$ are related by $\delta^2 \sim D \times \tau_{70}$, where $\delta$ is the typical distance over which low molecular weight species have to diffuse to reach the surface. The lengthscale $\delta$ is likely to be the thickness of the porous silica layer formed at the surface of the oxydized PDMS \cite{Kim2000}. Although it depends on parameters such as plasma intensity and exposure time \cite{Hillborg2001, Bayley2014, Sarrazin2016}, this thickness $\delta$ does not vary with temperature, hence the Arrhenius behavior of Eq. \eqref{eq:tau}, where $E_a$ can be interpreted as the diffusion activation energy of low molecular weight species in the porous silica layer.
%
\subsection{Wettability measurements in microchannels} \label{ssec:results_micro}
%
\begin{figure}[t]
\centering
\includegraphics[width=0.95\linewidth]{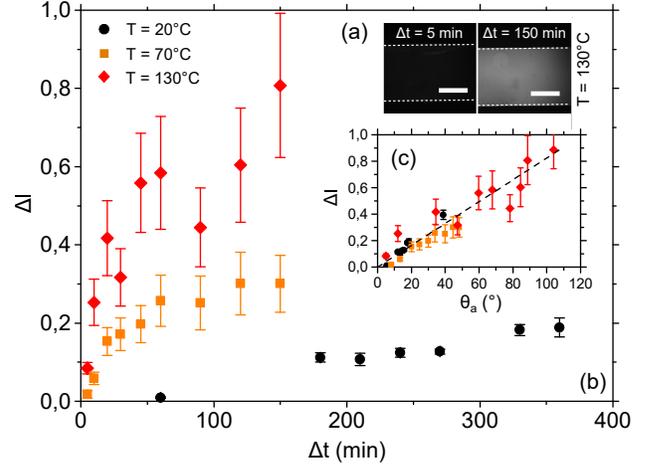}
\caption{(a) Top view of the microchannel illuminated by UV light after a $5~\minute$ and $150~\minute$ bake at $130~\celsius$ following the plasma treatment ($\Delta t=0$). The scale bars are $200~\micro\meter$. --- (b) Evolution of the average normalized fluorescence intensity $\Delta I$ in the channel (see Eq.~\eqref{eq:intensity}) as a function of the baking time $\Delta t$, when the sample is baked at room temperature ($20~\celsius$), $70$ and $130~\celsius$. Error bars represent the standard deviation of the fluorescence intensity. --- (c) Normalized fluorescence intensity $\Delta I$ in microchannels as a function of the corresponding contact angle $\theta_a$ measured on macroscopic PDMS surfaces baked in the same conditions. The dashed line is a linear fit of the data.}
\label{fig:micro}
\end{figure}
In this section, we present results corresponding to experiments of type (ii), where plasma-hydrophilized PDMS microchannels are further baked in ovens at various temperatures and for various amounts of time. The wettability in those channels is assessed by observing the fluorescence of BSA-FITC adsorbed onto the channel's walls, as this protein is known to preferentially adsorb on hydrophobic surfaces. Fluorescence pictures of a $400~\micro\meter$-wide PDMS microchannel are shown in Fig.~\ref{fig:micro}a, after a $5~\minute$ and a $150~\minute$ bake in an oven at $130~\celsius$ ($\Delta t=0$ corresponds to the time when the plasma treatment is performed). While no fluorescence -- hence no protein adsorption -- is visible on the freshly hydrophilized channel (left picture), a strong fluorescence signal is observed in the channel that has been heated for $150~\minute$ (right picture), demonstrating the adsorption of BSA-FITC on the hydrophobically-recovered PDMS walls.

The average intensity $I_{\mathrm{fluo}}$ of the fluorescence signal within the microchannel is systematically measured for various baking temperatures and baking times. This value is then corrected from the background intensity $I_0$ (average intensity outside of the microchannel) and normalized by the asymptotic value reached after a $24$-days hydrophobic recovery ($\Delta t \rightarrow \infty$) at the given temperature. The resulting normalized intensity $\Delta I$, defined as
\begin{equation}
\Delta I (\Delta t,T) = \frac{I_{\mathrm{fluo}}(\Delta t,T) - I_0(\Delta t,T)}{I_{\mathrm{fluo}}(\infty,T) - I_0(\infty,T)},
\label{eq:intensity}
\end{equation}
is plotted in Fig.~\ref{fig:micro}b as a function of time $\Delta t$ for baking temperatures of $20$, $70$ and $130~\celsius$. The fluorescence intensity in hydrophobically recovering microchannels is observed to follow the same qualitative trends as contact angle measurements on macroscopic PDMS surfaces. At a given temperature, $\Delta I$ increases with the baking time and the larger the temperature, the faster the recovery towards fully hydrophobic surfaces ($\Delta I = 1$). 

In order to evaluate how quantitative the fluorescence intensity measurements are, as a wettability probe, we correlate the contact angle measurements presented in Fig.~\ref{fig:macro} to the fluorescence intensity measurements. For each data point in Fig.~\ref{fig:micro}b, we compute the advancing contact angle on a macroscopic PDMS surface that has been baked in the same conditions (temperature and duration) as the microchannel, using the stretched exponential functions fitted in paragraph \ref{ssec:results_macro}. Fig.~\ref{fig:micro}c thus displays the fluorescence intensity $\Delta I$ observed within a microchannel as a function of the contact angle $\theta_a$ measured on the corresponding macroscopic PDMS surface.

A first important observation is that the intensities measured in microchannels baked at various temperatures all collapse onto the same curve when plotted as a function of the contact angle. This shows that fluorescence intensity does not depend on how a given contact angle was achieved (e.g. by baking the channel at a high temperature for a short time or at a low temperature for a long time).
A second observation is that the correlation between the fluorescence intensity $\Delta I$ and the contact angle is monotonous and approximately linear with $\theta_a$.
Both observations lead to the conclusion that the fluorescence intensity of BSA-FITC proteins adsorbed onto PDMS surfaces can be used as an indirect and quantitative indicator of the contact angle, and thus of the wettability inside PDMS microchannels. In our experimental conditions, $\Delta I \approx 0.00824 \times \theta_a $, where the contact angle is expressed in degrees.
Note that the normalization of the fluorescence intensity by its maximal value, obtained in the same conditions on a fully hydrophobic PDMS surface, is essential. The non-normalized intensity $I_{\mathrm{fluo}}$ may indeed depend on the concentration of the protein solution and the PBS flushing flow rate, but has also been observed to vary with the pH and ionic strength of the solution, as well as with the polarization of light \cite{Cheng1987}.

Finally, we briefly comment on the potential effects of using different plasma parameters for experiments (i) and (ii) (see paragraph \ref{ssec:PDMS}) on the calibration curve $\Delta I$ vs $\theta_a$ (Fig.~\ref{fig:micro}c). Following Bayley \textit{et al.}\cite{Bayley2014}, increasing the plasma power from $25~\watt$ (ii) to $100~\watt$ (i) in our experimental conditions increases the thickness of the surface oxidized layer from about $5$ to $20~\nano\meter$, thus slowing down the diffusion of low molecular weight chains to the surface. Concomitantly, the appearance of cracks in the oxydized layer at high plasma power\cite{Befahy2010} may instead accelerate this diffusion. Owing to these two antagonistic effects, we observed (data not shown) that the timescale of hydrophobic recovery changes only by a factor of $2$ upon decreasing the plasma power from $100~\watt$ to $25~\watt$. Shifting the contact angle data of Fig.~\ref{fig:macro} in time by the corresponding amount barely affects the calibration curve of Fig.~\ref{fig:micro}c, thereby showing the robustness of our conclusions.
%
\subsection{Wettability patterning in microchannels} \label{ssec:results_local}
%
\begin{figure}[t]
\centering
\includegraphics[width=1\linewidth]{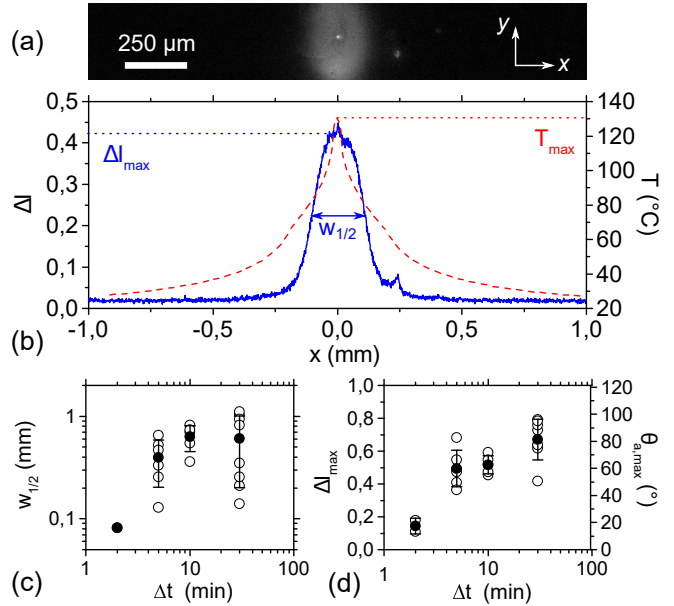}
\caption{(a) Fluorescence image of a microchannel heated locally by a $50~\micro\meter$ wide microresistor at a peak temperature $T_{\mathrm{max}}=130~\celsius$ during $\Delta t=30~\minute$. --- (b) Corresponding normalized intensity profile $\Delta I (x)$ as defined in Eq.~\eqref{eq:intensityX} (solid blue line) and temperature profile $T(x)$ generated by the resistor (dashed red line). --- Evolution of the intensity peak width at half-maximum $w_{1/2}$ (c) and maximum value $\Delta I_{\mathrm{max}}$ (d) as a function of the baking time $\Delta t$ using the temperature profile displayed in (b). Open circles are data obtained for several repetitions of the experiments and filled symbols show the corresponding average value and standard deviation.}
\label{fig:local}
\end{figure}
We now use the wettability probing technique described in the previous paragraph to demonstrate that hydrophobic regions can be patterned in a microchannel by locally increasing the temperature in the channel using a microresistor (see paragraph \ref{ssec:temp_control}).

As an example, Fig.~\ref{fig:local}a shows the fluorescence signal observed in a microchannel heated locally by a $50~\micro\meter$-wide microresistor at a peak temperature $T=130~\celsius$ during $\Delta t=30~\minute$.
Note that, because the temperature decreases sharply in the $y$-direction at the edges of the resistor (see Fig.~\ref{fig:resistors}c), the fluorescent intensity is only considered up to $y=\pm 150~\micro\meter$ from the microchannel central line.
The corresponding normalized intensity profile $\Delta I (x)$ (averaged across the $y$-direction) is displayed in Fig.~\ref{fig:local}b, along with the temperature profile $T(x)$ generated by the resistor. 
In this case, the definition of the normalized intensity $\Delta I$ (Eq.~\eqref{eq:intensity}) becomes a function of the space coordinate $x$ along the channel:
\begin{equation}
\Delta I (\Delta t, T(x)) = \frac{I_{\mathrm{fluo}}(\Delta t,T(x)) - I_0(\Delta t,T(x))}{I_{\mathrm{fluo}}(\infty,T_{\mathrm{max}}) - I_0(\infty,T_{\mathrm{max}})},
\label{eq:intensityX}
\end{equation}
where $I_{\mathrm{fluo}}(\Delta t,T(x))$ is the raw fluorescence intensity profile within the microchannel and $I_0(\Delta t,T(x))$ is the raw background intensity profile outside the microchannel. The normalization is done using the intensity $I_{\mathrm{fluo}}(\infty,T_{\mathrm{max}}) - I_0(\infty,T_{\mathrm{max}})$ obtained when the PDMS surface has fully recovered its hydrophobicity after having been heated at a temperature $T_{\mathrm{max}}$ corresponding to the peak value of the temperature profile.

Fig.~\ref{fig:local}a and b show that local heating of a PDMS surface with a stationary temperature profile $T(x)$ indeed gives rise to a fluorescence peak centered on the microresistor ($x\approx 0$) and characterized by its maximum value $\Delta I_{\mathrm{max}}$ and width at half-maximum $w_{1/2}$.
Repeating the experiment using the same temperature profile but various baking times, we obtain the width $w_{1/2}$ and amplitude $\Delta I_{\mathrm{max}}$ of the peak as a function of the baking time $\Delta t$, reported in Fig.~\ref{fig:local}c and d, respectively.
As established in paragraph \ref{ssec:results_micro}, this fluorescence enhancement corresponds to a local increase in hydrophobicity that can be quantified using the linear correspondence $\Delta I \propto \theta_a$.
Hence, in Fig.~\ref{fig:local}d, the maximum intensity value $\Delta I_{\mathrm{max}}$ of the fluorescence profile is converted into the corresponding maximum advancing contact angle $\theta_{a,\mathrm{max}}$.

Although the temperature profile $T(x)$ remains stationary, the wettability profile evolves with the baking time $\Delta t$ because low molecular weight chains diffuse much slower than heat in PDMS.
Both the amplitude $\Delta I_{\mathrm{max}}$ (or equivalently $\theta_{a,\mathrm{max}}$) and the width $w_{1/2}$ of the fluorescence peak exhibit increasing trends with $\Delta t$, as shown in Fig.~\ref{fig:local}c and d. 
Note that the widening of the fluorescence peak with time is most likely due to the lateral diffusion of low molecular weight species (\textit{i.e.} in directions parallel to the PDMS surface).
Despite the scattering of the measured peak widths $w_{1/2}$, one can estimate from Fig.~\ref{fig:local}c the typical resolution of the pattern to be a few hundreds of microns for a resistor width $w_x=50~\micro\meter$. 
From Fig.~\ref{fig:local}d, the time $\tau_{70}$ at which $\theta_{a,\mathrm{max}} = 70~\degree$ is found to be about $20 \pm 10~\minute$. 
This value is comparable to, albeit somewhat smaller than, the time $\tau_{70} = 60 \pm 5~\minute$ measured for PDMS samples submitted to a macroscopic heating at $T=T_{\mathrm{max}}=130~\celsius$ (see paragraph \ref{ssec:results_macro}).
The reason for this discrepancy may lie in the fact that PDMS elastomers dilate upon increasing temperature. Temperature gradients, such as those generated by our local heating device, cause a localized deformation of the PDMS \cite{Selva2010,Kerdraon2019} that may result in larger mechanical stresses as compared to when PDMS is heated (and therefore dilates) globally.
This may promote the apparition of cracks in the silica-like layer at the PDMS surface \cite{Fritz1995, Hillborg1998} and, as a consequence, accelerate the hydrophobic recovery of locally heated samples.
%
\section{Conclusions} \label{sec:conclusion}
%
The wettability of PDMS is a key parameter in many microfluidics applications, yet its control can be challenging. 
PDMS elastomers indeed naturally recover their original hydrophobicity after an oxidizing treatment, mainly because of the migration of low molecular weight chains towards the surface. 
Because this diffusion-controlled mechanism is strongly accelerated by temperature, we proposed to take advantage of hydrophobic recovery to pattern hydrophobic patches in otherwise hydrophilic microchannels by locally heating them. 
We first characterized the rate of hydrophobic recovery of PDMS macroscopic surfaces as a function of baking time and temperature, using (advancing) contact angle measurements. 
The same experiments were then carried out at microscale, using adsorption of a fluorescent protein (BSA-FITC) as a hydrophobicity marker in PDMS microchannels. 
Comparison between the two sets of data revealed a linear correlation between contact angle and fluorescence intensity in our experimental conditions.
This fluorescent wettability probe allowed us to characterize quantitatively the space-resolved hydrophobic patterns created in hydrophilized PDMS channels locally heated with a rectangular microresistor.

This ``thermo-patterning'' method could be easily extended to create more elaborated wettability patterns by changing the microresistor geometry and choosing adequate temperature and baking time. 
Because the local resistance depends on the local width of the resistor, smooth wettability gradients may also be obtained using resistors of varying width.
It should be stressed that the obtained wettability patterns are ephemeral, as the non-heated surfaces will slowly recover their original hydrophobicity with time. However, the durability of the pattern may be improved by simply storing the treated chips at low tempreature\cite{Bodin2017} or filling the channel with water or culture medium\cite{Zhao2012}.

One of the main interests of ``thermo-patterning'' is that hydrophobic patches of a few hundreds of microns in size can be patterned on demand in sealed PDMS chips, without need of any further chemical treatment. 
Space-resolved \textit{in situ} wettability patterning techniques indeed usually require advanced polymer chemistry\cite{Hu2004, Abate2008, Schneider2010, Romanowsky2010}. Additionally, simpler methods to pattern sealed chips, using for example spatially controlled oxydation\cite{Kim2015} or flow confinement of chemical treatement\cite{Abate2010}, can only be applied to entire channels. 
Because it simply takes advantage of the natural polydispersity of commercial PDMS elastomers, our technique does not raise any biocompatibility issues. 
``Thermo-patterning'' therefore stands as a valuable complement to existing \textit{in situ} wettability patterning techniques.
%
\section*{Acknowledgments}
The authors are grateful to St\'ephanie Descroix for introducing them to Bovine Serum Albumine and to Javier Rodr\'iguez Rodr\'iguez for fruitful discussions regarding the manuscript.

This work was supported by CNRS,  ESPCI Paris, Agence Nationale de la Recherche (ANR) under the grant 13-BS09-0011-01, IPGG (Equipex ANR-10-EQPX-34 and Labex ANR-10-LABX-31), PSL (Idex ANR-10-IDEX-0001-02) and Dim-NanoK R\'{e}gion Ile de France.

%
\bibliography{biblio}
\bibliographystyle{ieeetr}
%

\end{document}